\documentclass[11pt]{article} 
\newcommand{\be}{\begin{equation}}
\newcommand{\ee}{\end{equation}}
\newcommand{\bea}{\begin{eqnarray}}
\newcommand{\eea}{\end{eqnarray}}
\newcommand{\sn}{{\rm sn}}

\newcommand{\dn}{{\rm dn}}
\newcommand{\cn}{{\rm cn}}
\newcommand{\sech}{{\rm sech}}

\begin{document}
\vspace{.5in} 
\begin{center} 
{\LARGE{\bf Novel PT-invariant Kink  and Pulse Solutions For a Large 
Number of  Real Nonlinear Equations}}
\end{center} 

\vspace{.3in}
\begin{center} 
{\LARGE{\bf Avinash Khare}} \\ 
{Physics Department, Savitribai Phule Pune University \\
Pune, India 411007}
\end{center} 

\begin{center} 
{\LARGE{\bf Avadh Saxena}} \\ 
{Theoretical Division and Center for Nonlinear Studies, Los
Alamos National Laboratory, Los Alamos, NM 87545, USA}
\end{center} 

\vspace{.9in}
{\bf {Abstract:}}  

For a large number of real nonlinear equations, either continuous 
or discrete, integrable or nonintegrable, uncoupled or coupled, 
we show that whenever a real nonlinear equation admits kink solutions 
in terms of $\tanh \beta x$, where $\beta$ is the inverse of the kink width, 
it also admits solutions in terms of the PT-invariant combinations 
$\tanh 2\beta x \pm i \sech 2 \beta x$, i.e. the kink width is reduced by half
to that of the real kink solution. We show that both the kink and the 
PT-invariant 
kink are linearly stable and obtain expressions for the zero mode in
the case of several PT-invariant kink solutions. Further, 
for a number of real nonlinear equations we show that whenever a nonlinear 
equation admits periodic kink solutions in terms of $\sn(x,m)$, it also admits 
periodic solutions in terms of the PT-invariant combinations 
$\sn(x,m) \pm i \cn(x,m)$ 
as well as $\sn(x,m)\pm i \dn(x,m)$. Finally, for coupled equations
we show that one cannot only have complex PT-invariant solutions 
with PT eigenvalue $+1$ or $-1$ in both the fields but one can also have
solutions with PT eigenvalue $+1$ in one field and $-1$ in the other field.

\newpage 
  
\section{Introduction} 

Nonlinear equations are playing an increasingly important role in several areas 
of science in general and physics in particular \cite{book}. 
One of the major problems with 
these equations is the lack of a superposition principle. 
In particular, even if one can find two solutions, say $\phi_1$ and 
$\phi_2$, of a given nonlinear equation, unlike the linear case, any
linear combination of $\phi_1$ and $\phi_2$ is usually not a solution
of that nonlinear equation. Thus if we can find some general 
results about the existence of solutions to nonlinear equations, that would 
be invaluable. In this context it is worth 
recalling that some time ago we \cite{ks1,ks2} have shown (through a number of
examples) that if a nonlinear
equation admits periodic solutions in terms of Jacobi elliptic functions 
$\dn(x,m)$ and $\cn(x,m)$, then it will also admit solutions in terms 
of $\dn(x,m) \pm \sqrt{m} \cn(x,m)$ where m is the modulus of the 
elliptic function \cite{as}. 
Further, in the same papers \cite{ks1,ks2}, we also showed (again through 
several examples) that if a 
nonlinear equation admits solutions in terms of $\dn^2(x,m)$, then it will 
also admit solutions in terms of $\dn^2(x,m) \pm \cn(x,m) \dn(x,m)$.

The purpose of this paper is to propose general results about the existence 
of new solutions to real nonlinear equations, integrable or nonintegrable, 
continuous or discrete through the idea of parity-time reversal or PT symmetry. 
It may be noted here that in the last 15 years or so the idea of PT symmetry 
\cite{ben} has given us new insights. In quantum mechanics it has been shown 
that even if a Hamiltonian is not hermitian but if it is PT-invariant, then the energy 
eigenvalues are still real in case the PT symmetry is not broken spontaneously. 
Further, there has been a tremendous growth in the number of studies of open 
systems which are specially balanced by PT symmetry \cite{sch,ben1,peng} in 
several PT-invariant open systems bearing both loss and gain. In particular, 
many researchers have obtained soliton solutions which have been shown 
to be stable within a certain parameter range \cite{mak,pan,jes}. 

It is worth specifying what exactly we mean by $P$ and $T$ and hence the 
PT symmetry. By $P$ one means parity symmetry, i.e. $x \rightarrow -x$, $t
\rightarrow t$ while by $T$ one means time-reversal symmetry, i.e.
$t \rightarrow -t$, $i \rightarrow -i$, $x \rightarrow x$. Thus by the
combined PT symmetry we mean $x \rightarrow -x$, $t \rightarrow -t$, 
$i \rightarrow -i$.

In this context, recently we have highlighted one more novel aspect of 
PT symmetry.  Specifically, 
we obtained new PT-invariant solutions of several real nonlinear equations with
PT-eigenvalue +$1$. In particular, we showed \cite{ks} that if a real 
nonlinear equation admits soliton
solutions in terms of $\sech x$ then it also admits PT-invariant solutions
$\sech x \pm i \tanh x$ with PT-eigenvalue $+1$.  We also showed that if 
a real nonlinear equation admits
solutions in terms of $\sech^2 x$ then it also admits PT-invariant solutions
$\sech^2 x \pm i \sech x \tanh x$ with PT-eigenvalue $+1$. 
In addition, we considered the periodic
generalization of these results. It is worth pointing out that in all these
cases the PT-invariant combinations (such as $\sech x \pm i \tanh x$) 
are eigenfunctions of the PT operator with eigenvalue $+1$.  

It is then natural to inquire if there are PT-invariant
solutions with PT-eigenvalue $-1$ and further in the case of coupled field
theories, are there PT-invariant solutions with PT-eigenvalue $-1$ in
both the fields and also if there are mixed PT-invariant solutions with 
PT-eigenvalue $+1$ in one field and $-1$ in the other field. One of the 
aims of this paper is to provide answers to these questions. Our 
strategy will be to start with known real solutions and then make Ans\"atze 
for complex PT-invariant solutions and obtain conditions under which the 
Ans\"atze are valid. We show, through several examples (such as $\phi^4$, 
$\phi^6$, sine-Gordon, double sine-Gordon, double sine-hyperbolic-Gordon 
equations), that whenever a real nonlinear
equation, either continuous or discrete, integrable or nonintegrable, 
admits a solution in terms of $\tanh \beta x$, then it will necessarily
also admit solutions in terms of the PT-invariant combinations
$\tanh 2\beta x \pm i \sech 2\beta x$ with PT-eigenvalue $-1$ (i.e. with
the PT-kink width being half of  that of the corresponding real kink). 
Remarkably, in all these cases, the kink solution as well as the 
PT-invariant kink are solutions of the first order self-dual equation.
Further in all these cases, the kink as well as the PT-invariant kink
have the same topological charge and the same kink energy and both 
the solutions can be shown to be linearly stable. In view of this, we believe 
that the PT-invariant kink solutions may also find physical realization in 
coupled optical waveguides among other applications \cite{ruter}.
It is worth pointing out that some of the equations considered here have also
been considered in their PT-symmetric deformed version \cite{beb}.

We also generalize these 
results to the periodic case and show that whenever a nonlinear equation
admits a solution in terms of $\sn(x,m)$, then it will 
necessarily also admit solutions in terms of the PT-invariant combinations 
$\sn (x,m)\pm i\cn (x,m)$  as well as  $\sn(x,m) \pm i \dn(x,m)$ with 
PT-eigenvalue $-1$.  Further, we also consider coupled field theory models 
and obtain PT-invariant solutions with PT-eigenvalue $-1$ in both the fields 
(in addition to PT-eigenvalue $+1$ in both the fields) and mixed solutions
with PT-eigenvalue $+1$ in one field and $-1$ in the other field. 

The plan of the paper is the following. In Sec. II we consider several
self-dual first order equations which are known to admit topological
kink solutions of the form $\tanh \beta x$ and in all these cases show the 
existence of PT-invariant complex kink solutions of the form $\tanh 2\beta x 
\pm i \sech 2\beta x$ with PT-eigenvalue $-1$ and kink width being half of that 
of the corresponding real kink.  We show that a given kink solution and the 
corresponding PT-invariant kink solution have the same topological charge, 
the same kink energy and further, both are linearly stable. For all the 
PT-invariant kink solutions we give explicit expressions for the zero mode. 
In Sec. III we consider four continuum and two discrete field theory models 
and show the existence of PT-invariant periodic kink solutions of the form 
$\sn(x,m) \pm i \cn(x,m)$ and $\sn(x,m) \pm i \dn(x,m)$ with 
PT-eigenvalue $-1$ and also the corresponding hyperbolic PT-invariant kink
solutions. In Sec. IV  we discuss three coupled field theory models
and show that these models not only admit PT-invariant solutions with
PT-eigenvalue $+1$ or $-1$ for both the fields, but also mixed PT-invariant
solutions with PT-eigenvalue $+1$ in one field and $-1$ in the other field.
Section V is reserved for summary of the main results obtained where we also 
discuss some of the open problems.

\section{PT-invariant Kink Solutions} 

We now discuss several examples from continuum field theories 
where a kink solution like $\tanh \beta x$ is a  
solution of the first order self-dual equation. We consider several 
self-dual first order equations with known kink solutions and in all 
the cases obtain new PT-invariant solutions in terms of 
$\tanh 2\beta x \pm i \sech 2\beta x$ with PT-eigenvalue $-1$. 

As is well known, typically, the self-dual equations with kink
solutions are of the form
\be\label{2.0}
\frac{d\phi}{dx} = \pm \sqrt{2V(\phi)}\,,
\ee
where $V(\phi)$ has multiple degenerate minima and is positive semidefinite 
with its minimum value being zero at the degenerate minima.
We show in general that  
both the kink topological charge as well as the kink energy remain 
unaltered, i.e. the usual kink solution and the new PT-invariant kink
solutions have the same topological charge and the same kink energy.
Further, we show that both the usual kink as well as the PT-invariant kink
solutions are linearly stable and give explicit expressions
for the nodeless zero mode in all the cases.

\subsection{$\phi^4$ Kink}

The $\phi^4$ field theory arises in several areas of physics.
It has the celebrated kink solution which is the solution of the 
first order self-dual equation
\be\label{1.14}
\frac{d\phi}{dx} = \pm \sqrt{\frac{b}{2}}\left(\frac{a}{b}-\phi^2\right)\,.
\ee
We consider here and in rest of the examples, one of the self-dual equations. 
Exactly the same arguments are also valid for the other cases.
It is well known that the kink solution to Eq. (\ref{1.14}) is 
\cite{raj}
\be\label{1.15}
\phi_{k} = A \tanh \beta x\,,
\ee
provided
\be\label{1.16}
A = \sqrt{\frac{a}{b}}\,,~~\beta = \sqrt{\frac{a}{2}}\,.
\ee
The corresponding topological charge is 
\be\label{1.17}
Q = \phi (x=\infty) - \phi (x=-\infty)= 2\sqrt{\frac{a}{b}}\,,
\ee
while the corresponding kink energy is 
\bea\label{1.18}
&&E_k = \int_{-\infty}^{+\infty} dx\, \left[\frac{1}{2}\left(\frac{d\phi}{dx}\right)^2 +V(\phi)\right]
\nonumber \\
&&= \int_{-\sqrt{\frac{a}{b}}}^{\sqrt{\frac{a}{b}}} d\phi\, \sqrt{2V(\phi)} 
= \frac{4}{3}\sqrt{\frac{a^3}{b^3}}\,.
\eea

Remarkably, even 
\be\label{1.19}
\phi_{ptk} = A \tanh \beta_1 x +i B\sech \beta_1 x\,,
\ee
is an exact PT-invariant kink solution (with PT eigenvalue $-1$) 
of the self-dual Eq. (\ref{1.14}) provided
\be\label{1.20}
B = \pm A\,,~~ A = \sqrt{\frac{a}{b}}\,,~~ \beta_1 = \sqrt{2a} = 2 \beta\,,
\ee
where $\beta$ is as given by Eq. (\ref{1.16}).
Note that the corresponding topological charge as defined by Eq. (\ref{1.17}) 
is the same for the
kink solution (\ref{1.15}) and the complex PT-invariant kink 
solution (\ref{1.19}). This is because  
$\phi(\pm \infty)$ remains unchanged. Further, even the kink 
energy is the same for the kink solution (\ref{1.15}) and the PT-invariant 
kink solution (\ref{1.19}) since as is clear from Eq. (\ref{1.18}),
the answer for the kink energy depends on $V(\phi)$ and $\phi(\pm \infty)$
both of which  are the same for the kink solution (\ref{1.15}) as well 
for the PT-invariant kink solution (\ref{1.19}).

The arguments given above are rather general and hence it is clear that 
the topological charge and the kink energy are the same for any (real) 
kink solution
and the corresponding complex PT-invariant kink solution. We have also
explicitly checked it for all the cases mentioned below. Hence now onwards, 
for the remaining examples, we shall not discuss the kink topological 
charge and the kink energy for the PT-invariant kink solutions.   

Let us now discuss the question of linear stability of the complex
PT-invariant kink solution (\ref{1.19}). In this context it is worth 
recalling that
in the case of the standard kink solution as given by Eq. (\ref{1.15}),
the linear stability issue has been discussed a while ago \cite{raj}
and it has been shown that on assuming
\be\label{1.14a}
\phi = \phi_{k} + \eta(x) e^{i\omega t}\,,
\ee
and substituting it in the corresponding (second-order) field equation, 
one gets the stability equation
\be\label{1.15a}
-\frac{d^2 \eta(x)}{dx^2} + \frac{d^2 V(\phi_{k})}{dx^2} \eta(x) 
= \omega^2 \eta(x)\,.
\ee
Further, it is well known that the corresponding zero mode, i.e. 
the unnormalized ground state wave function $\eta_0$
with $\omega^2 =0$ is given by
\be\label{1.15d}
\eta_{0}(x) = \frac{d\phi_{k}}{dx}\,.
\ee
This discussion is easily generalized to the PT-invariant kink and
in that case one gets the stability equation which is similar to 
(\ref{1.15a}) except that $\frac{d^2V(\phi_{k})}{dx^2}$ is
replaced by $\frac{d^2V(\phi_{ptk})}{dx^2}$ and the corresponding
zero-mode is given by
\be\label{1.15e}
\eta_{0,pt}(x) = \frac{d\phi_{ptk}}{dx}\,.
\ee

For the usual $\phi^4$ kink, using Eq. (\ref{1.15}), one gets the 
stability equation
\be\label{1.16a}
-\frac{d^2\eta(x)}{dx^2}+[2a-3a \sech^2(\beta x)]\eta(x) = \omega^2 \eta(x)\,,
~~\beta = \sqrt{\frac{a}{2}}\,.
\ee
As is well known, this Schr\"odinger-like equation has two bound states with
the corresponding eigenvalues and eigenfunctions being
\bea\label{1.17a}
&&\eta_{0}(x) = \sech^2 (\beta x)\,,~~ \omega^{2}_{0} = 0\,,
\nonumber \\
&&\eta_{1}(x) = \sech(\beta x) \tanh(\beta x)\,,~~\omega^{2}_{1} 
= \frac{3a}{2}\,.
\eea
Note that while $\eta_{0}$ is nodeless, $\eta_{1}$ has one node.

Let us now discuss the stability of the PT-invariant kink solution 
(\ref{1.19}). On using Eq. (\ref{1.19}) in the stability 
Eq. (\ref{1.15a}), one gets a Schr\"odinger-like equation for the 
PT-invariant potential
\be\label{1.18a}
-\frac{d^2\eta(x)}{dx^2}+[2a-6a \sech^2(\beta_1 x) 
\pm 6ia \sech(\beta_1 x) \tanh(\beta_1 x)]\eta(x) = \omega^2 \eta(x)\,,
~~\beta_1 = \sqrt{2a}\,.
\ee
Remarkably, this Schrodinger-like PT-invariant equation too has two 
discrete states but unlike the real kink case, both the discrete states
are nodeless but with different energies. In particular, the 
corresponding eigenvalues and eigenfunctions are
\bea\label{1.19a}
&&\eta_{0,pt}(x) = \sech (\beta_1 x)[\sech(\beta_1 x)\mp i \tanh(\beta_1 x)]
= \sech(\beta_1 x) e^{\mp i\tan^{-1}[\sinh(\beta_1 x)]}\,,
~~ \omega_{0} = 0\,,
\nonumber \\
&&\eta_{1,pt}(x) = \sech^{1/2}(\beta_1 x) e^{\mp (3i/2)
\tan^{-1}[\sinh(\beta_1 x)]} \,,~~\omega^{2}_{1} 
= \frac{3a}{2}\,.
\eea
It is worth noting that the eigenenergies of the two bound states are
identical (i.e. $0, \sqrt{3a/2}$) for the usual and the PT-invariant 
kink solution. 

Because of the linear stability of the PT-invariant kink solution, 
we suspect that the PT-invariant kink solution too can have physical 
relevance. It would be interesting to explore physical situations where
such kink can indeed be relevant.

We now show that similar discussion is valid in the case of the other 
PT-invariant kink solutions and in all the cases the zero mode is nodeless
thereby ensuring the linear stability of the PT-invariant kink solution.

\subsection{$\phi^6$ Kinks}

Unlike the $\phi^4$ case, in the $\phi^6$ case there are two different
types of kink solutions depending on if one is at the first order 
transition point or below the transition point. We now show that in both
the cases we have PT-invariant kink solutions.

{\bf Case I: $T = T_{c1}$}

In this case the kink solution satisfies the self dual equation  
\be\label{1.21}
\frac{d\phi}{dx} = \pm \sqrt{\lambda} \phi (a^2-\phi^2)\,.
\ee
At this point there are two distinct kink solutions of the form 
$\sqrt{1\pm \tanh \beta x}$  and in both the cases the 
PT-invariant kink
solutions also exist. For simplicity we only discuss one case, 
exactly the same arguments are also valid in the other case.

One of the well known kink solutions to Eq. (\ref{1.21}) 
is \cite{kh}
\be\label{1.22}
\phi_{k} = A \sqrt{1+\tanh \beta x}\,,
\ee
provided
\be\label{1.23}
A = a/\sqrt{2}\,,~~\beta = \sqrt{\lambda} a^2\,.
\ee
The self-dual Eq. (\ref{1.21}) also admits the PT-invariant
kink solution 
\be\label{1.24}
\phi_{ptk} = A \sqrt{1+\tanh \beta_1 x \pm i \sech \beta_1 x}\,,
\ee
provided $A$ is again as given by Eq. (\ref{1.23}) while the inverse 
width $\beta$ is again doubled, i.e. 
\be\label{1.25}
\beta_1 = 2\sqrt{\lambda} a^2 = 2 \beta\,,
\ee
where $\beta$ is given by Eq. (\ref{1.23}).

Let us now show that this PT-invariant kink solution is linearly stable. 
From Eq. (\ref{1.21}) it is clear that
\be\label{1.21a}
V(\phi) = \frac{\lambda}{2}\phi^2(a^2-\phi^2)^2\,.
\ee
It is then easy to calculate $\frac{d^2V(\phi_{ptk})}{d\phi^2}$ and set up
the stability equation for the PT-invariant kink solution (\ref{1.24}). 
In particular, on setting $y = \beta_1 x$, the stability Eq. (\ref{1.15a})
takes the form
\be\label{1.21f}
-\eta''(y)+(1/8)[5-15 \sech^2 y+3\tanh y\pm 3i\sech y 
\pm 15 i \sech y \tanh y]\eta(y) = \frac{\omega^2}{4\lambda a^4} \eta (y)\,.
\ee
On using Eq. (\ref{1.15e}) the corresponding zero-mode turns out to be
\be\label{1.23a}
\eta_{0,pt} \propto (\sech y)^{1/2}(1-\tanh y)^{1/4}
e^{\mp(3i/4)\tan^{-1}[\sinh y]}\,,~~\omega_{0} =0\,.
\ee

{\bf Case II: $T < T_{c1}$}

In this case the self-dual equation is of the form 
\be\label{1.26}
\frac{d\phi}{dx} = \pm \sqrt{\lambda} (a^2-\phi^2)\sqrt{\phi^2+b^2}\,.
\ee
The well known kink solution to this equation is \cite{chle,ks4} 
\be\label{1.27a}
\phi = \frac{A\tanh \beta x}{\sqrt{1-B^2 \tanh^2 \beta x}}\,,
\ee
provided
\be\label{1.28}
A = \frac{ab}{\sqrt{a^2+b^2}}\,,~~B = \frac{a^2}{a^2+b^2}\,,~~
\beta = a \sqrt{\lambda (a^2+b^2)}\,.
\ee
The solution (\ref{1.27a}) can be put in the form
\be\label{1.27}
\frac{\phi}{\sqrt{A^2+B\phi^2}} = \tanh \beta x\,,
\ee
The self-dual Eq. (\ref{1.26}) also admits the PT-invariant
kink solution
\be\label{1.29}
\frac{\phi}{\sqrt{A^2+B\phi^2}} = \tanh \beta_1 x \pm i \sech \beta_1 x\,,
\ee
provided $A$ and $B$ are as given by Eq. (\ref{1.28}) while $\beta$ is
again doubled and given by
\be\label{1.30}
\beta_1 = 2a\sqrt{\lambda(a^2+b^2)} = 2\beta\,,
\ee
where $\beta$ is given by Eq. (\ref{1.28}).

Let us now show that this PT-invariant kink solution is linearly stable. 
From Eq. (\ref{1.26}) it is clear that
\be\label{1.26a}
V(\phi) = \frac{\lambda}{2}(b^2+\phi^2)(a^2-\phi^2)^2\,.
\ee
It is then easy to calculate $\frac{d^2V(\phi_{ptk})}{d\phi^2}$ and set up
the stability equation for the PT-invariant kink solution (\ref{1.29}).
In particular, on putting $y = \beta_1 x$, the stability Eq. (\ref{1.15a})
takes the form 
\bea\label{1.26f}
&&-\eta''(y)+\frac{1}{4(a^2+b^2)} \Bigg(a^2-2b^2 
+\frac{3b^2(b^2-2a^2)}{2a^2}\left[\frac{(b^2/(b^2+a^2)-2\sech^2 y 
\pm 2i \sech y \tanh y}{\sech^2 y + p^2}\right] \nonumber \\
&&+\frac{15b^4}{16a^2}\left[\frac{(b^2/(b^2+a^2)-2\sech^2 y 
\pm 2i \sech y \tanh y}{\sech^2 y + p^2}\right]^2 \Bigg) \eta(y) 
= \frac{\omega^2}{4a^2 \lambda (a^2+b^2)}\eta(y)\,,
\eea
where $p^2 = \frac{b^4}{4a^2(a^2+b^2)}$. 
On using Eq. (\ref{1.15e}), we obtain the well known zero-mode for the 
PT-invariant kink solution (\ref{1.29}) 
\bea\label{1.28a}
&&\eta_{0,pt} \propto \sech y
\sqrt{\frac{b^2}{a^2+b^2}-2\sech^2 y \mp 2i \sech y \tanh y} \nonumber \\
&& \times \frac{[b^2/(a^2+b^2)](1+\frac{b^2}{2a^2}) \sech y \tanh y 
\pm i(1+2p^2)\sech^2 y \mp i p^2}{[\sech^2 y+p^2]^2}\,.
\eea

\subsection{Sine-Gordon Kink}

The self-dual sine-Gordon equation is \cite{raj}
\be\label{1.31}
\frac{d\phi}{dx} = \pm 2 \sin \frac{\phi}{2}\,.
\ee
One of the well known kink solutions is  
\be\label{1.32a}
\phi = 4 \tan^{-1}(e^{-x})\,,  
\ee
which can be put in the form
\be\label{1.32}
\cos(\phi/2) = \tanh \beta x\,,~~\beta =1\,.
\ee
The self-dual Eq. (\ref{1.31}) also admits the PT-invariant
kink solution
\be\label{1.33}
\cos(\phi/2) = \tanh \beta_1 x \pm i \sech \beta_1 x\,,
\ee
provided $\beta$ is again doubled, i.e. $\beta_1 =2 = 2\beta$ where
$\beta$ is given by Eq. (\ref{1.32}).

Let us now show that this PT-invariant kink solution is linearly stable. 
From Eq. (\ref{1.31}) it is clear that
\be\label{1.31a}
V(\phi) = 1- \cos(\phi)\,.
\ee
It is then easy to calculate $\frac{d^2V(\phi_{ptk})}{d\phi^2}$ and set up
the stability equation for the PT-invariant kink solution (\ref{1.33}). 
In particular, on substituting $y = \beta_1 x$ in Eq. (\ref{1.15a})
we obtain the stability equation
\be\label{1.31f}
-\eta''(y)+\left[\frac{1}{4} - \sech^2 y \pm i \sech y \tanh y\right]\eta(y) = 
\frac{\omega^2}{4} \eta(y)\,.
\ee
Using Eq. (\ref{1.15e}), the zero-mode
for the PT-invariant kink solution (\ref{1.33}) turns out to be
\be\label{1.33a}
\eta_{0,pt} \propto \sqrt{\sech y}\, e^{\mp (i/2)\tan^{-1}(\sinh y)}\,.
\ee

\subsection{Double sine-hyperbolic-Gordon Kink}

The self-dual equation for the double sine-hyperbolic-Gordon (DSHG) equation is
\be\label{1.34}
\frac{d\phi}{dx} = \pm \sqrt{2}(\zeta \cosh 2\phi -n)\,.
\ee
The well known kink solution in this case is \cite{raz,bk,hks}
\be\label{1.35}
\phi = \tanh^{-1}\left[\sqrt{\frac{n-\zeta}{n+\zeta}}\tanh \beta x\right]\,,~~\beta =
\sqrt{n^2-\zeta^2}\,,
\ee
 which can be put in the form
\be\label{1.36}
\tanh \phi = \sqrt{\frac{n-\zeta}{n+\zeta}} \tanh \beta x\,.
\ee
The same self-dual Eq. (\ref{1.34}) also admits the PT-invariant
kink solution
\be\label{1.37}
\tan \phi = \sqrt{\frac{n-\zeta}{n+\zeta}}\, 
[\tanh \beta_1 x \pm i\sech \beta_1 x]\,,
\ee
provided $\beta$ is doubled, i.e. $\beta_1 = 2\sqrt{n^2-\zeta^2} = 2\beta$
where $\beta$ is given by Eq. (\ref{1.35}). 

Let us now show that this PT-invariant kink solution is linearly stable. 
From Eq. (\ref{1.34}) it is clear that
\be\label{1.31d}
V(\phi) = (\zeta \cosh 2\phi -n)^2\,.
\ee
It is then easy to calculate $\frac{d^2V(\phi_{ptk})}{d\phi^2}$ and set up
the stability equation for the PT-invariant kink solution (\ref{1.37}). 
In particular, on substituting $y = \beta_1 x$ in Eq. (\ref{1.15a})
we obtain the stability equation
\bea\label{1.37f}
&&-\eta''(y)+\frac{1}{n-\zeta}\bigg (2\zeta -\frac{2\zeta}
{(n^2 \sech^2 y+\zeta^2 \tanh^2 y)}[2\zeta 
+(n+4\zeta)(n\sech^2 y+\zeta \tanh^2 y \pm i(n-\zeta)\sech 6 \tanh y]
\nonumber \\
&&+\frac{8\zeta^2(n+\zeta)(n\sech^2 y + \zeta \tanh^2 y)}
{(n^2\sech^2 y+\zeta^2 \tanh^2 y)^2}[n\sech^2 y+\zeta \tanh^2 y 
\pm i (n-\zeta)\sech y \tanh y] \bigg ) \eta (y) = 
\frac{\omega^2}{4(n^2-\zeta^2)} \eta(y)\,. \nonumber \\
\eea
Using Eq. (\ref{1.15e}), the zero-mode
for the PT-invariant kink solution (\ref{1.37}) turns out to be
\be\label{1.33d}
\eta_{0,pt} \propto \frac{\sech y (n\sech y \mp i \zeta \tanh y)}
{n^2 \sech^2 y+\zeta^2 \tanh^2 y}\,. 
\ee

\subsection{Double Sine-Gordon Kink}

Consider the following self-dual equation for the double sine-Gordon case
\be\label{1.38}
\frac{d\phi}{dx} = \pm \sqrt{2\lambda} \left(\sin \phi - \frac{\mu}{\lambda}\right)\,,
~~\mu < \lambda\,.
\ee
In this case the well known kink solution is \cite{leung}
\be\label{1.39a}
\phi = 2\tan^{-1}\left(\sqrt{\frac{\lambda - \mu}{\lambda + \mu}} 
\tanh \beta x\right) +\frac{\pi}{2}\,,
\ee
which can be put in the form
\be\label{1.39}
\tan\left(\frac{\phi}{2}- \frac{\pi}{4}\right) = \sqrt{\frac{\lambda - \mu}
{\lambda + \mu}}\tanh(\beta x)\,,
\ee
provided
\be\label{1.40}
\beta = \sqrt{\frac{\lambda(1-\frac{\mu^2}{\lambda^2})}{2}}\,.
\ee
The same self-dual Eq. (\ref{1.38}) admits the PT-invariant
kink solution
\be\label{1.41}
\tan\left(\frac{\phi}{2}- \frac{\pi}{4}\right) = \sqrt{\frac{\lambda - \mu}
{\lambda + \mu}}\,[\tanh \beta x \pm i \sech \beta x]\,,
\ee
provided $\beta$ is doubled, i.e.
\be\label{1.42}
\beta_1 = \sqrt{2\lambda\left(1-\frac{\mu^2}{\lambda^2}\right)} = 2\beta\,,
\ee
where $\beta$ is given by Eq.  (\ref{1.40}).

Let us now show that this PT-invariant kink solution is linearly stable. 
From Eq. (\ref{1.38}) it is clear that
\be\label{1.31g}
V(\phi) = \lambda \left(\sin \phi - \frac{\mu}{\lambda}\right)^2\,.
\ee
It is then easy to calculate $\frac{d^2V(\phi_{ptk})}{d\phi^2}$ and set up
the stability equation for the PT-invariant kink solution (\ref{1.41}). 
In particular, on substituting $y = \beta_1 x$ in Eq. (\ref{1.15a})
we obtain the stability equation
\bea\label{1.41f}
&&-\eta''(y)+\frac{\lambda^2}{\lambda^2-\mu^2} \bigg [1
+\frac{\mu(\mu \lambda \mp i (\lambda^2 - \mu^2)\sech y}
{\lambda(\mu^2 \sech^2 y+\lambda^2 \tanh^2 y)} \nonumber \\
&&-2\frac{\mu^2 \lambda^2 -(\lambda^2-\mu^2)^2 \sech^2 y 
\mp 2i\mu \lambda (\lambda^2-\mu^2) \sech y}
{(\mu^2 \sech^2 y + \lambda^2 \tanh^2 y)^2} \bigg ]\eta(y) =
\frac{\omega^2 \lambda}{2(\lambda^2-\mu^2)}\eta(y)\,.
\eea
Using Eq. (\ref{1.15e}), the zero-mode
for the PT-invariant kink solution (\ref{1.41}) turns out to be
\be\label{1.41a}
\eta_{0,pt} \propto \frac{\sech y (\mu \sech y \mp i \lambda \tanh y)}
{\mu^2 \sech^2 y+\lambda^2 \tanh^2 y}\,. 
\ee

\section{PT-Invariant Periodic Kink Solutions}

We now discuss a few examples from both the continuum and the discrete 
field theories where both periodic and hyperbolic kink-like solutions 
are known, and show that in all these cases one also has complex 
PT-invariant periodic as well as hyperbolic kink solutions.

\subsection{mKdV Equation}

We first discuss the celebrated mKdV equation
\be\label{2.1}
u_t+u_{xxx}- 6 u^2 u_{x} =0\,,
\ee
which is a well known integrable equation having application in several 
areas \cite{dj}.
One of the exact, periodic solutions to the mKdV Eq. (\ref{2.1}) is 
\be\label{2.2}
u = A \sqrt{m} \sn[\beta(x-vt),m]\,,
\ee
provided
\be\label{2.3}
A^2 =  \beta^2\,,~~v= -(1+m) \beta^2\,.
\ee
In the limit $m=1$, the solution (\ref{2.2}) goes over to the hyperbolic 
kink solution
\be\label{2.4}
u = A \tanh[\beta(x-vt)\,,
\ee
and in this case $v = -2\beta^2$.

Remarkably, even the complex PT-invariant combination (with PT eigenvalue
$-1$) 
\be\label{2.5}
u = A\sqrt{m} \sn[\beta(x-vt),m] + i B \sqrt{m} \cn[\beta(x-vt),m]\,,
\ee
is an exact solution to the mKdV Eq. (\ref{2.1}) provided
\be\label{2.6}
B = \pm A\,,~~\beta^2 = 4 A^2\,,~~v= -\frac{(2-m)}{2}\beta^2\,.
\ee

Yet another PT-invariant solution (with PT eigenvalue $-1$) is
\be\label{2.7}
u = A \sqrt{m} \sn[\beta(x-vt),m] +i B \dn[\beta(x-vt),m]\,,
\ee
provided
\be\label{2.8}
B = \pm A\,,~~\beta^2 = 4 A^2\,,~~v = -\frac{(2m-1)}{2} \beta^2\,.
\ee
We thus have two new periodic solutions of the mKdV Eq. (\ref{2.1}). 
In the limit $m=1$, both these solutions go over to the hyperbolic 
PT-invariant solution
\be\label{2.9}
u = A \tanh[\beta(x-vt)] \pm i B \sech[\beta(x-vt)]\,,
\ee
provided
\be\label{2.10}
B = \pm A\,,~~\beta^2 = 4 A^2\,,~~v= -(1/2) \beta^2\,  . 
\ee

There is also a complex PT-invariant solution to the 
mKdV Eq. (\ref{2.1}) with PT-eigenvalue +$1$. Let us first note that
the mKdV Eq. (\ref{2.1}) has an exact solution
\be\label{2.11}
u = \frac{A\sqrt{m}\cn[\beta(x-vt),m]}{\dn[\beta(x-vt),m]}\,,
\ee
provided
\be\label{2.12}
A^2 = \beta^2\,,~~v = -(1+m)\beta^2\, .
\ee
It is easily checked that the same Eq. (\ref{2.1}) also has the complex
PT invariant solution with PT-eigenvalue +$1$
\be\label{2.13}
u = A \sqrt{m}\,\frac{\cn[\beta(x-vt),m]}{\dn[\beta(x-vt),m]} 
+iB \sqrt{m(1-m)} \frac{\sn[\beta(x-vt),m]}{\dn[\beta(x-vt),m]}\,,
\ee
provided
\be\label{2.14c}
B = \pm A\,,~~\beta^2 = 4A^2\,,~~v = -\frac{(2-m)}{2}\beta^2\,.
\ee

Before completing this subsection, we would like to note that in our recent
paper \cite{ks} we had considered the other mKdV equation, i.e.
\be\label{2.15}
u_t+u_{xxx}+ 6 u^2 u_{x} =0\,,
\ee
and had shown that in that case one has complex PT-invariant solutions of the 
form $\cn(x,m) \pm i\sn(x,m)$ and $\dn(x,m) \pm i\sn(x,m)$ with PT-eigenvalue 
$+1$. We now show that Eq. (\ref{2.15})  also has a PT-invariant solution 
with PT-eigenvalue $-1$. Let us first note that
\be\label{2.16}
u = A \sqrt{m(1-m)} \frac{\sn[\beta(x-vt),m]}{\dn[\beta(x-vt),m]}\,,
\ee
is an exact solution to Eq. (\ref{2.15}) provided
\be\label{2.17}
A^2 = \beta^2\,,~~v = (2m-1)\beta^2\, .
\ee
It is easily checked that the same Eq. (\ref{2.15}) also has the complex
PT invariant solution with PT-eigenvalue $-1$
\be\label{2.18}
u = A \sqrt{m(1-m)} \frac{\sn[\beta(x-vt),m]}{\dn[\beta(x-vt),m]}\,,
+iB \sqrt{m} \frac{\cn[\beta(x-vt),m]}{\dn[\beta(x-vt),m]}\,,
\ee
provided
\be\label{2.14}
B = \pm A\,,~~4A^2 = \beta^2\,,~~v = -\frac{(2-m)}{2}\beta^2\,.
\ee

\subsection{$\phi^4$ Field Theory}

The field equation in this case is
\be\label{2.20}
\phi_{xx} = a \phi + b \phi^3\,,
\ee
which also follows from the self-dual first order Eq. (\ref{1.14}). 
We now show that in this case too one has complex PT-invariant solutions
with PT eigenvalue $-1$. 

One of the exact, periodic solutions to the $\phi^4$ Eq. (\ref{2.20}) 
is \cite{aubry} 
\be\label{2.21}
u = A \sqrt{m} \sn(\beta x,m)\,,
\ee
provided
\be\label{2.22}
b A^2 =  2\beta^2\,,~~a= -(1+m) \beta^2\,.
\ee
In the limit $m=1$, the solution (\ref{2.21}) goes over to the hyperbolic 
kink solution discussed in the previous section.

Remarkably, even the complex PT-invariant combination (with PT eigenvalue
$-1$) 
\be\label{2.23}
u = A\sqrt{m} \sn(\beta x,m) + i B \sqrt{m} \cn(\beta x,m)\,,
\ee
is an exact solution to Eq. (\ref{2.20}) provided
\be\label{2.24}
B = \pm A\,,~~\beta^2 = 2b A^2\,,~~a= -\frac{(2m-1)}{2}\beta^2\,.
\ee

Yet another PT-invariant solution (with PT eigenvalue $-1$) is
\be\label{2.25}
u = A \sqrt{m} \sn(\beta x,m) +i B \dn(\beta x,m)\,,
\ee
provided
\be\label{2.26}
B = \pm A\,,~~\beta^2 = 2b A^2\,,~~a = -\frac{(2-m)}{2} \beta^2\,.
\ee
In the limit $m=1$, both solutions (\ref{2.23}) and (\ref{2.25}) go
over to the hyperbolic PT-invariant kink solution discussed in the 
previous section.

Another periodic solution to Eq. (\ref{2.20}) is
\be\label{2.27}
\phi = A \sqrt{m(1-m)}\frac{\sn(\beta x,m)}{\dn(\beta x,m)}\,,
\ee
provided
\be\label{2.28}
b A^2 = - 2\beta^2\,,~~a = -(2m-1)\beta^2\,.
\ee 
The complex PT-invariant combination (with PT-eigenvalue $-1$)
\be\label{2.29}
\phi = A \sqrt{m(1-m)}\frac{\sn(\beta x,m)}{\dn(\beta x,m)}
+iB \sqrt{m} \frac{\cn(\beta x,m)}{\dn(\beta x,m)}\,,
\ee
is also an exact solution to Eq. (\ref{2.20}) provided
\be\label{2.30}
B = \pm A\,,~~2b A^2 = -\beta^2\,,~~a = - \frac{(2-m)}{2} \beta^2\,.
\ee

So far we have discussed complex PT-invariant solutions (with PT-eigenvalue
$-1$) of the $\phi^4$ field Eq. (\ref{2.20}). Further, in  a recent paper
we have already obtained complex PT-invariant periodic solutions of the
$\phi^4$ field Eq. (\ref{2.20}) with PT-eigenvalue $+1$. They were of the
form $\cn(x,m) \pm i\sn(x,m)$ and $\dn(x,m) \pm i\sn(x,m)$. We now show 
that the same Eq. (\ref{2.20}) also has another periodic PT-invariant solution 
with PT-eigenvalue $+1$. Let us first note that one of the exact solutions to
Eq. (\ref{2.20}) is
\be\label{2.31}
\phi = A \sqrt{m}\frac{\cn(\beta x,m)}{\dn(\beta x,m)}\,,
\ee
provided
\be\label{2.32}
b A^2 =  2\beta^2\,,~~a = -(1+m)\beta^2\,.
\ee 
The complex PT-invariant combination (with PT-eigenvalue $+1$)
\be\label{2.33}
\phi = A \sqrt{m}\frac{\cn(\beta x,m)}{\dn(\beta x,m)}
+iB \sqrt{m(1-m)} \frac{\sn(\beta x,m)}{\dn(\beta x,m)}\,,
\ee
is also an exact solution to Eq. (\ref{2.20}) provided
\be\label{2.34}
B = \pm A\,,~~2b A^2 = \beta^2\,,~~a = - \frac{2-m}{2} \beta^2\,.
\ee

\subsection{KdV Equation}

In our recent publication \cite{ks} we have also obtained complex PT-invariant 
solutions of the KdV equation
\be\label{2.60}
u_t+u_{xxx}- 6 u u_{x} =0\,,
\ee
with PT eigenvalue +$1$. We now discuss one more complex PT-invariant
periodic solution of this equation with PT-eigenvalue $+1$.
To begin with, it is easy to check that one of the exact periodic solutions
of the KdV Eq. (\ref{2.60}) is
\be\label{2.61}
u = \frac{A(1-m)}{\dn^2[\beta(x-vt),m]}\,,
\ee
provided
\be\label{2.61a}
A =  -2\beta^2\,,~~v= 4 (2-m) \beta^2\,.
\ee
The same equation also admits the complex, PT-invariant, periodic
solution  
\be\label{2.62}
u = \frac{A(1-m)}{\dn^2[\beta(x-vt),m]} +i B \sqrt{1-m}  
\frac{m\sn[\beta(x-vt),m] \cn[\beta(x-vt),m]}{\dn^2[\beta(x-vt),m]}\,,
\ee
provided
\be\label{2.63}
B = \pm A\,,~~A = -\beta^2\,,~~v= (2-m)\beta^2\,.
\ee
It may be noted that (\ref{2.62}) is a nonsingular, periodic solution
which vanishes in the hyperbolic limit $m=1$.

\subsection{$\phi^3$ Field Theory}

This field theory arises in the context of third order phase transitions 
\cite{phi3} and is also relevant to tachyon condensation \cite{tachyon}.
In our recent publication \cite {ks} we also discussed complex PT-invariant
periodic solutions of the $\phi^3$ field theory with PT-eigenvalue $+1$.
In this subsection, we discuss one more complex, PT-invariant, periodic
solution with PT-eigenvalue $+1$. We first notice that one of the exact
solutions of the $\phi^3$ field theory
\be\label{2.64}
\phi_{xx} = a\phi+b\phi^2\,,
\ee
is
\be\label{2.65}
\phi = \frac{A(1-m)}{\dn^2[\beta(x),m]}+ Ay\,,
\ee
provided
\be\label{2.66}
bA = -6\beta^2\,,~~a=4[2-m+3y]\beta^2\,,~~y = 
\frac{-(2-m)\pm \sqrt{1-m+m^2}}{3}\,.
\ee
The same Eq. (\ref{2.64}) also admits the complex, PT-invariant
periodic solution (with PT-eigenvalue $+1$)
\be\label{2.67}
\phi = \frac{A(1-m)}{\dn^2[\beta(x),m]}+A y+ i B m\sqrt{1-m} 
\frac{\cn[\beta(x),m] \sn[\beta(x),m]}{\dn^2[\beta x,m]}\,,
\ee
provided
\be\label{2.68}
B = \pm A\,,~~bA = -3\beta^2\,,~~a = (2-m+6y)\beta^2\,,~~
y = \frac{-(2-m)\pm \sqrt{16-16m+m^2}}{6}\,.
\ee
Note that this is a nonsingular, complex, PT-invariant solution which
vanishes in the hyperbolic limit $m = 1$.

\subsection{Discrete $\phi^4$ Equation}

We now discuss two discrete models and show that both these models also admit
PT-invariant periodic and hyperbolic kink solutions. Let us consider 
the discrete $\phi^4$ equation
\be\label{2.40}
\frac{1}{h^2}[\phi_{n+1}+\phi_{n-1}-2\phi_n]+ a \phi_n 
-\frac{\lambda}{2} \phi_{n}^{2}[\phi_{n+1}+\phi_{n-1}] = 0\,.
\ee

It is easy to check that Eq. (\ref{2.40}) admits an exact periodic
kink solution \cite{cooper}
\be\label{2.41}
\phi_n = A \sqrt{m} \sn(\beta n,m)\,,
\ee
provided
\be\label{2.42}
A^2  =  \frac{2 \sn^2(\beta,m)}{h^2 \lambda}\,,~~
a h^2 = 2[1-\cn(\beta,m) \dn(\beta,m)]\,.
\ee

The same model also admits a PT-invariant periodic kink solution
\be\label{2.43}
\phi_n = A\sqrt{m} \sn(\beta n,m) +i B\sqrt{m} \cn(\beta n,m)\,,
\ee
provided
\be\label{2.44}
B = \pm A\,,~~A^2   = \frac{2\sn^2 (\beta,m)}{h^2 \lambda [1+\dn(\beta,m)]^2}
\,,~~a h^2 = 2\left[1-\frac{2\cn(\beta,m)}{1+\dn(\beta,m)}\right]\,.
\ee

Further, the model also admits yet another PT-invariant
periodic kink solution
\be\label{2.45}
\phi_n = A\sqrt{m} \sn(\beta n,m) +i B \dn(\beta n,m)\,,
\ee
provided
\be\label{2.46}
B = \pm A\,,~~A^2   = \frac{2\sn^2 (\beta,m)}{h^2 \lambda [1+\cn(\beta,m)]^2}
\,,~~a h^2 = 2\left[1-\frac{2\dn(\beta,m)}{1+\cn(\beta,m)}\right]\,.
\ee

In the limit $m=1$, both the solutions (\ref{2.43}) and (\ref{2.45}) go over
to the hyperbolic PT-invariant kink solution
\be\label{2.47}
\phi_n = A \tanh(\beta n) +i B \sech(\beta n)\,,
\ee
provided
\be\label{2.48}
B = \pm A\,,~~A^2   = \frac{2\tanh^2 (\frac{\beta}{2})}
{h^2 \lambda}\,,~~a h^2 = 2\tanh^2\left(\frac{\beta}{2}\right)\,.
\ee

While deriving results in this and the next subsection, we have made use of
several not so well known identities satisfied by the Jacobi elliptic 
functions \cite{kls}.

\subsection{Discrete mKdV Equation}

Let us consider the discrete mKdV equation
\be\label{2.49}
\frac{du_n}{dt}+  \alpha (u_{n+1} -u_{n-1})
-\lambda u_{n}^{2}(u_{n+1} -u_{n-1}) =0\,. 
\ee
It is easily checked that this model admits the periodic kink solution 
\cite{dmkdv}
\be\label{2.50}
u_n = A\sqrt{m} \sn[\beta (n-vt),m]\,,
\ee
provided
\be\label{2.51}
\lambda A^2 = \alpha \sn^2(\beta, m)\,,
~~\beta v = 2\alpha \sn(\beta,m)\,.
\ee

Remarkably, the same model (\ref{2.49}) also admits a complex PT-invariant
periodic kink solution
\be\label{2.52}
u_n = A \sqrt{m} \sn[\beta (n-vt),m] +iB \sqrt{m} \cn[\beta (n-vt),m]\,,
\ee
provided
\be\label{2.53}
B = \pm A\,,~~\lambda A^2 = \frac{\alpha \sn^2(\beta, m)}
{[1+\dn(\beta,m)]^2}\,,~~\beta v = \frac{4\alpha \sn(\beta,m)}
{1+\dn(\beta,m)}\,.
\ee
Further, the same model also admits yet another complex PT-invariant 
periodic kink solution
\be\label{2.54}
u_n = A \sqrt{m} \sn[\beta (n-vt),m] +iB \dn[\beta (n-vt),m]\,,
\ee
provided
\be\label{2.55}
B = \pm A\,,~~\lambda A^2 = \frac{\alpha \sn^2(\beta, m)}
{[1+\cn(\beta,m)]^2}\,,~~\beta v = \frac{4\alpha \sn(\beta,m)}
{1+\cn(\beta,m)}\,.
\ee

In the limit $m=1$, both the complex PT-invariant solutions
(\ref{2.52}) and (\ref{2.54}) go over to the complex PT-invariant
hyperbolic kink solution
\be\label{2.56}
u_n = A \tanh(\beta n) + i B\sech(\beta n)\,,
\ee
provided
\be\label{2.57}
B = \pm A\,,~~\lambda A^2 = \alpha \tanh^2(\beta/2)\,,
~~\beta v  = 4 \alpha \tanh(\beta/2) \,.
\ee

\section{PT-Invariant Solutions in Three Coupled  models}

We now consider three coupled models and show that in all these cases
one has PT-invariant solutions for the coupled fields. In particular,
we show that these models admit three different types of complex, 
PT-invariant periodic as well as hyperbolic solutions. In particular, 
there are solutions with PT 
eigenvalue $+1$ or $-1$ in both the fields and also solutions with
PT eigenvalue $+1$ in one field and $-1$ in the other field. 

\subsection{Coupled $\phi^4$ Model}

We first consider a coupled $\phi^4$ model 
\be\label{3.1}
\phi_{xx} = a_1 \phi +  b_1 \phi^3 + \alpha \phi \psi^2\,,
\ee
\be\label{3.2}
\psi_{xx} = a_2 \psi +  b_2 \psi^3 + \alpha \psi \phi^2\,,
\ee
and show that in this case all three types (i.e. those with PT-eigenvalue
$+1$ or $-1$ in both the fields or $+1$ in one field and $-1$ in the other
field) of PT-invariant periodic as well as hyperbolic solutions are allowed.
We shall see that there are solutions not only in terms of Lam\'e polynomials
of order one but also in terms of  Lam\'e polynomials of order two.

\subsubsection{Solutions in Terms of Lam\'e Polynomials of order one}

Let us first discuss solutions in terms of Lam\'e polynomials of order 
one. 

{\bf Case I: Solutions With PT Eigenvalue $-1$ in Both The Fields}

We first discuss PT-invariant solutions with PT eigenvalue $-1$
in both the fields.

It is easy to check that one of the exact solutions to Eq. (\ref{3.1}) is
\cite{ks3}
\be\label{3.3}
\phi = A\sqrt{m} \sn[\beta x,m]\,,~~\psi = B \sqrt{m} \sn[\beta x, m]\,,
\ee
provided
\be\label{3.4}
b_1 A^2 +\alpha B^2 = b_2 B^2 + \alpha A^2 = 2\beta^2\,,
~~a_1 = a_2 = -(1+m)\beta^2\,.
\ee

The same coupled model also admits the PT-invariant periodic solution
\bea\label{3.5}
&&\phi = A \sqrt{m} \sn[\beta x,m]+ i D\sqrt{m} \cn[\beta x,m]\,, \nonumber \\
&&\psi = B \sqrt{m} \sn[\beta x,m]+ i F\sqrt{m} \cn[\beta x,m]\,,
\eea
provided
\be\label{3.6}
D = \pm A\,,~~F = \pm B\,,~~a_1 = a_2 
= -\frac{(2-m)\beta^2}{2}\,,
\ee 
and further
\be\label{3.7}
2(b_1 A^2+\alpha B^2) = 2(b_2 B^2+\alpha A^2) = \beta^2\,.
\ee
Note that the signs of $D = \pm A$ and $F = \pm B$ are correlated.

Further, the same model also admits another PT-invariant periodic solution
\bea\label{3.8}
&&\phi = A \sqrt{m} \sn[\beta x,m]+ i D \dn[\beta x,m]\,, \nonumber \\
&&\psi = B \sqrt{m} \sn[\beta x,m]+ i F  \dn[\beta x,m]\,,
\eea
provided
\be\label{3.9}
D = \pm A\,,~~F = \pm B\,,~~a_1 = a_2 
= -\frac{(2m-1)\beta^2}{2}\,,
\ee
and if Eq. (\ref{3.7}) is satisfied.
Note that the signs of $D = \pm A$ and $F = \pm B$ are correlated.

In the limit $m=1$, both the periodic PT-invariant solutions (\ref{3.5})
and (\ref{3.8}) go over to the coupled hyperbolic PT-invariant solution
\bea\label{3.10}
&&\phi = A \tanh(\beta x)+ i D \sech(\beta x)\,, \nonumber \\
&&\psi = B \tanh(\beta x)+ i F \sech(\beta x)\,,
\eea
provided Eq. (\ref{3.7}) is satisfied and if further
\be\label{3.11}
D = \pm A\,,~~F = \pm B\,,~~a_1 = a_2 
= -\frac{\beta^2}{2}\,.
\ee
On solving Eq. (\ref{3.7}) we find that
\be\label{3.12}
A^2 = \frac{(b_2 -\alpha)\beta^2}{2(b_1 b_2 -\alpha^2)}\,,
~~B^2 = \frac{(b_1 -\alpha)\beta^2}{2(b_1 b_2 -\alpha^2)}\,,
\ee
provided $b_1 b_2 \ne \alpha^2$. In the special case when $b_1 b_2 = \alpha^2$
which implies $b_1 = b_2 = \alpha$, instead of the two relations of 
Eq. (\ref{3.7}), we only have a single relation
\be\label{3.13}
2b_1 (A^2+B^2) = \beta^2\,,
\ee
and hence, in this case, $A,B$ are indeterminate except that they 
satisfy the constraint (\ref{3.13}). 

Yet another exact solution to  Eqs. (\ref{3.1}), (\ref{3.2}) is
\bea\label{3.3a}
&&\phi = A\sqrt{m(1-m)} \frac{\sn(\beta x, m)}{\dn(\beta x, m)}\,,
\nonumber \\
&&\psi = B\sqrt{m(1-m)} \frac{\sn(\beta x, m)}{\dn(\beta x, m)}\,,
\eea
provided
\be\label{3.4a}
b_1 A^2 +\alpha B^2 = b_2 B^2 + \alpha A^2 = -2\beta^2\,,
~~a_1 = a_2 = (2m-1)\beta^2\,.
\ee

Remarkably, we find that the same coupled model also admits the PT-invariant
periodic solution 
\bea\label{3.5a}
&&\phi = A \sqrt{m(1-m)} \frac{\sn(\beta x, m)}{\dn(\beta x, m)}
+ i D\sqrt{m} \frac{\cn(\beta x, m)}{\dn(\beta x, m)}\,, \nonumber \\
&&\psi = B \sqrt{m(1-m)} \frac{\sn(\beta x, m)}{\dn(\beta x, m)}
+ i F\sqrt{m} \frac{\cn(\beta x, m)}{\dn(\beta x, m)}\,, 
\eea
with PT-eigenvalue $-1$ in both the fields provided
\be\label{3.6a}
D = \pm A\,,~~F = \pm B\,,~~a_1 = a_2 
= -\frac{(2-m)\beta^2}{2}\,,
\ee 
and further
\be\label{3.7a}
2(b_1 A^2+\alpha B^2) = 2(b_2 B^2+\alpha A^2) = -\beta^2\,.
\ee
Note that the signs of $D = \pm A$ and $F = \pm B$ are correlated.
On solving Eq. (\ref{3.7a}) we find that
\be\label{3.12a}
A^2 = -\frac{(b_2 -\alpha)\beta^2}{2(b_1 b_2 -\alpha^2)}\,,
~~B^2 = -\frac{(b_1 -\alpha)\beta^2}{2(b_1 b_2 -\alpha^2)}\,,
\ee
provided $b_1 b_2 \ne \alpha^2$. In the special case when $b_1 b_2 = \alpha^2$
which implies $b_1 = b_2 = \alpha$, instead of the two relations of 
Eq. (\ref{3.7a}), we only have a single relation
\be\label{3.13a}
2b_1 (A^2+B^2) = -\beta^2\,,
\ee
and hence, in this case, $A,B$ are indeterminate except that they 
satisfy the constraint (\ref{3.13a}). 

\vskip 0.1truein
\noindent{\bf Case II: Solutions with Mixed PT Eigenvalues}

We now discuss mixed PT-invariant solutions, i.e.
PT-invariant solutions  with PT eigenvalue +$1$ in one field and $-1$ 
in the other field.

It is easy to check that one of the exact solutions to Eqs. (\ref{3.1}), (\ref{3.2}) is
\be\label{3.14}
\phi = A\sqrt{m} \sn[\beta x,m]\,,~~\psi = B \sqrt{m} \cn[\beta x, m]\,,
\ee
provided
\be\label{3.15}
b_1 A^2 -\alpha B^2 = \alpha A^2 -b_2 B^2 = 2\beta^2\,,
~~a_1 +m \alpha B^2 = -(1+m)\beta^2\,,
~~a_2 +m \alpha A^2 = (2m-1)\beta^2\, . 
\ee

We find that the same coupled model also admits the mixed 
PT-invariant periodic solution
\bea\label{3.16}
&&\phi = A \sqrt{m} \sn[\beta x,m]+ i D\sqrt{m} \cn[\beta x,m]\,, \nonumber \\
&&\psi = B \sqrt{m} \cn[\beta x,m]+ i F\sqrt{m} \sn[\beta x,m]\,,
\eea
provided
\be\label{3.17}
D = \pm A\,,~~F = \pm B\,,~~a_1 = a_2 
= -\frac{(2-m)\beta^2}{2}\,,
\ee 
and further
\be\label{3.18}
2(b_1 A^2-\alpha B^2) = 2(\alpha A^2 -b_2 B^2) = \beta^2\,.
\ee
Note that the signs of $D = \pm A$ and $F = \pm B$ are correlated.

It is easy to check that one of the exact solutions to 
Eqs. (\ref{3.1}), (\ref{3.2}) is
\be\label{3.14a}
\phi = A\sqrt{m} \sn[\beta x,m]\,,~~\psi = B \dn[\beta x, m]\,,
\ee
provided
\be\label{3.15a}
b_1 A^2 -\alpha B^2 = \alpha A^2 -b_2 B^2 = 2\beta^2\,,
~~a_1 + \alpha B^2 = -(1+m)\beta^2\,,
~~a_2 + \alpha A^2 = (2-m)\beta^2\, .
\ee

The same model also admits another PT-invariant periodic solution
\bea\label{3.19}
&&\phi = A \sqrt{m} \sn[\beta x,m]+ i D \dn[\beta x,m]\,, \nonumber \\
&&\psi = B \dn[\beta x,m]+ i F \sqrt{m} \sn[\beta x,m]\,,
\eea
provided
\be\label{3.20}
D = \pm A\,,~~F = \pm B\,,~~a_1 = a_2 
= -\frac{(2m-1)\beta^2}{2}\,,
\ee
and further if Eq. (\ref{3.18}) is satisfied.
Note that the signs of $D = \pm A$ and $F = \pm B$ are correlated.

In the limit $m=1$, both the periodic PT-invariant solutions (\ref{3.17})
and (\ref{3.19}) go over to the coupled, hyperbolic, mixed
 PT-invariant solution
\bea\label{3.21}
&&\phi = A \tanh(\beta x)+ i D \sech(\beta x)\,, \nonumber \\
&&\psi = B \sech(\beta x)+ i F \tanh(\beta x)\,,
\eea
provided Eq. (\ref{3.18}) is satisfied and if further
\be\label{3.22}
D = \pm A\,,~~F = \pm B\,,~~a_1 = a_2 
= -\frac{\beta^2}{2}\,.
\ee
On solving Eq. (\ref{3.18}) we find that
\be\label{3.23}
A^2 = \frac{(b_2 -\alpha)\beta^2}{2(b_1 b_2 -\alpha^2)}\,, ~~~
B^2 = \frac{(\alpha-b_1)\beta^2}{2(b_1 b_2 -\alpha^2)}\,,
\ee
provided $b_1 b_2 \ne \alpha^2$. In the special case when $b_1 b_2 = \alpha^2$
which implies $b_1 = b_2 = \alpha$, instead of the two relations of 
Eq. (\ref{3.18}), we only have a single relation
\be\label{3.24}
2b_1 (A^2-B^2) = \beta^2\,.
\ee
Thus $A,B$ are indeterminate in this case except that they must satisfy 
the constraint (\ref{3.24}). 

Yet another periodic solution to the coupled Eqs. (\ref{3.1}), (\ref{3.2}) is
\bea\label{3.3b}
&&\phi = A\sqrt{m} \frac{\cn(\beta x, m)}{\dn(\beta x, m)}\,,
\nonumber \\
&&\psi = B\sqrt{m(1-m)} \frac{\sn(\beta x, m)}{\dn(\beta x, m)}\,,
\eea
provided Eq. (\ref{3.15}) is satisfied.

We find that the same coupled model also admits the PT-invariant
periodic solution
\bea\label{3.5b}
&&\phi = A \sqrt{m} \frac{\cn(\beta x, m)}{\dn(\beta x, m)}
+ i D\sqrt{m(1-m)} \frac{\sn(\beta x, m)}{\dn(\beta x, m)}\,, \nonumber \\
&&\psi = B \sqrt{m(1-m)} \frac{\sn(\beta x, m)}{\dn(\beta x, m)}
+ i F\sqrt{m} \frac{\cn(\beta x, m)}{\dn(\beta x, m)}\,, \nonumber \\
\eea
provided Eq. (\ref{3.18}) is satisfied and if further
\be\label{3.6b}
D = \pm A\,,~~F = \mp B\,,~~a_1 = a_2 = -\frac{(2-m)\beta^2}{2}\,.
\ee 
Note that the signs of $D = \pm A$ and $F = \mp B$ are correlated.

\vskip 0.1truein 
\noindent{\bf Case III: Solutions With PT Eigenvalue +$1$ in Both The Fields}

We now discuss complex PT-invariant periodic solutions
with PT-eigenvalue $+1$ in both the fields. In our earlier paper we have
already discussed PT-invariant solutions of the form $\cn(x,m) \pm i\sn(x,m)$ 
and $\dn(x,m) \pm i \sn(x,m)$ with PT-eigenvalue $+1$.  We now show that 
there is another PT-invariant solution with PT-eigenvalue $+1$ in both the fields.

One exact solution to  Eqs. (\ref{3.1}), (\ref{3.2}) is
\be\label{3.25}
\phi = A\sqrt{m} \frac{\cn(\beta x, m)}{\dn(\beta x, m)}\,,
~~\psi = B\sqrt{m} \frac{\cn(\beta x, m)}{\dn(\beta x, m)}\,,
\ee
provided Eq. (\ref{3.4}) is satisfied.
We find that the same coupled model also admits the PT-invariant
periodic solution with PT-eigenvalue $+1$ in both the fields
\bea\label{3.27}
&&\phi = A \sqrt{m} \frac{\cn(\beta x, m)}{\dn(\beta x, m)}
+ i D\sqrt{m(1-m)} \frac{\sn(\beta x, m)}{\dn(\beta x, m)}\,, \nonumber \\
&&\psi = B \sqrt{m} \frac{\cn(\beta x, m)}{\dn(\beta x, m)}
+ i F\sqrt{m(1-m)} \frac{\sn(\beta x, m)}{\dn(\beta x, m)}\,, \nonumber \\
\eea
provided Eqs. (\ref{3.6}) and (\ref{3.7}) are satisfied.

\subsubsection{Solutions in Terms of Lam\'e Polynomials of Order Two}

We now show that for the coupled $\phi^4$ model (\ref{3.1}), (\ref{3.2}) one has
 all three types (i.e. those with PT-eigenvalue
$+1$ or $-1$ for both the fields or $+1$ in one field and $-1$ in the other
field) of PT-invariant solutions are allowed in terms of Lam\'e polynomials
of order two.

\vskip 0.1truein
\noindent {\bf Case I: Solutions With PT Eigenvalue $-1$ in Both The Fields}

It is easy to check that
\be\label{3.29}
\phi = A m \frac{\cn(\beta x, m)\sn(\beta x, m)}{\dn^2(\beta x, m}\,,
~~\psi = B \sqrt{m(1-m)} \frac{\sn(\beta x, m)}{\dn^2(\beta x, m}\,,
\ee
is an exact solution to the coupled Eqs. (\ref{3.1}), (\ref{3.2}) provided
\be\label{3.30}
b_1 = b_2 = \alpha\,,~~B = \pm A\,,~~b_1 A^2 = -6(1-m)\beta^2\,,~~
a_1 = (5m-4)\beta^2\,,~~a_2 = (5m-1)\beta^2\,.
\ee
Remarkably, the PT-invariant combination with PT-eigenvalue $-1$
\bea\label{3.31}
&&\phi = A m \frac{\cn(\beta x, m)\sn(\beta x, m)}{\dn^2(\beta x, m}
+iD\left[\frac{1-m}{\dn^2(\beta x, m)}+y\right] \nonumber \\
&&\psi = B \sqrt{m(1-m)} \frac{\sn(\beta x, m)}{\dn^2(\beta x, m}
+iF\sqrt{m} \frac{\cn(\beta x, m)}{\dn^2(\beta x, m}\,,
\eea
is also an exact solution to Eqs. (\ref{3.1}), (\ref{3.2}) provided
\be\label{3.32}
b_1 = b_2 = \alpha\,,~~D = \pm A\,,~~F = \mp B\,,~~2 b_1 A^2 y = 3 \beta^2\,,
\ee
\be\label{3.34}
a_1 = b_1 A^2 +[2-m+(9/2)y]\beta^2\,,~~
a_2 = b_1 A^2 +[2-m+(3/2)y]\beta^2\,,
\ee
and $y$ is given by Eq. (\ref{2.68}).

\vskip 0.1truein
\noindent{\bf Case II: Solutions With Mixed PT Eigenvalues}

Now let us discuss the so called mixed PT-invariant solutions, i.e. those with
PT-eigenvalue $+1$ in one field and $-1$ in the other field. 

It is easy to check that
\be\label{3.35}
\phi = A [\dn^2(\beta x, m)+y]\,,~~\psi = B m \sn(\beta x, m) \cn(\beta x, m)\,,
\ee
is an exact solution to the coupled Eqs. (\ref{3.1}), (\ref{3.2}) provided
\be\label{3.36}
b_1 = b_2 = \alpha\,,~~B = \pm A\,,~~(2y+2-m) b_1 A^2 = -6 \beta^2\,,
\ee
\be\label{3.37}
a_1 = [4(2-m)+6y]\beta^2 -[y^2-(1-m)]b_1 A^2\,,~~
a_2 = (2-m)\beta^2 -[y^2-(1-m)]b_1 A^2\,,
\ee
and $y$ is given by Eq. (\ref{2.66}).

There is a related PT-invariant solution with PT-eigenvalue  $-1$
in one field and +$1$ in the other field. In particular,
\bea\label{3.39}
&&\phi = A [\dn^2(\beta x, m)+y] +iD m \cn(\beta x, m) \sn(\beta x, m)\,,
\nonumber \\
&&\psi = B m \cn(\beta x, m) \sn(\beta x, m) +iF [\dn^2(\beta x, m)+z]\,,
\eea
is an exact solution to Eqs. (\ref{3.1}), (\ref{3.2}) provided
\be\label{3.40}
b_1 = b_2 = \alpha\,,~~B = \pm A\,,~~F = \mp B\,,~~ y \ne z\,,~~
(y-z) b_1 A^2 y = -(3/2) \beta^2\,,
\ee
\be\label{3.41}
a_1 = [2-m+(3/2)(3y+z)]\beta^2\,,~~
a_2 = [2-m+(3/2)(3z+y)]\beta^2\, , 
\ee
and both $y$ and $z$ are different; they are given by the two roots of Eq. (\ref{2.68}).

There is another PT-invariant solution with PT-eigenvalue $-1$
in one field and $+1$ in the other field which is related to the solution
(\ref{3.35}). In particular,
\bea\label{3.39a}
&&\phi = A [\dn^2(\beta x, m)+y] +iD \sqrt{m} \dn(\beta x, m) \sn(\beta x, m)\,,
\nonumber \\
&&\psi = B m \cn(\beta x, m) \sn(\beta x, m)
+iF \sqrt{m} \cn(\beta x,m) \dn(\beta x, m)\,,
\eea
is an exact solution to Eqs. (\ref{3.1}), (\ref{3.2}) provided
\be\label{3.40a}
b_1 = b_2 = \alpha\,,~~D = \pm A\,,~~F = \mp B\,,~~
(y+1-m) b_1 A^2  = -(3/2) \beta^2\,,
\ee
\be\label{3.41a}
a_1 = [5-4m+(9/2)y]\beta^2+(1-m)(y+1)b_1 A^2\,,~~
a_2 = [2-m+(3/2)y]\beta^2 +(1-m)(y+1)b_1 A^2\,,
\ee
and $y$ is given by
\be\label{3.42b}
y = \frac{-(5-4m)\pm\sqrt{1-16m+16 m^2}}{6}\, . 
\ee

In the limit $m=1$, both the solutions (\ref{3.39}) and (\ref{3.39a}) 
go over to the corresponding hyperbolic PT-invariant solution
\bea\label{3.39d}
&&\phi = A [\sech^2(\beta x)+y] +iD \sech(\beta x) \tanh(\beta x)\,,
\nonumber \\
&&\psi = B  \sech(\beta x) \tanh(\beta x)
+iF \sech^2(\beta x)\,,
\eea
provided
\be\label{3.40d}
b_1 = b_2 = \alpha\,,~~D = \pm A\,,~~F = \mp B\,,~~
y= -1/3,~~ z=0\,,~~b_1 A^2  = (9/2) \beta^2\,,
\ee
\be\label{3.41d}
a_1 = -(1/2)\beta^2\,,~~a_2 = (1/2)\beta^2\,.
\ee

It is easy to check that
\be\label{3.43}
\phi = A [\dn^2(\beta x, m)+y]\,,~~\psi = B \sqrt{m} \sn(\beta x, m) 
\dn(\beta x, m)\,,
\ee
is an exact solution to the coupled Eqs. (\ref{3.1}), (\ref{3.2}) provided
\be\label{3.44}
b_1 = b_2 = \alpha\,,~~B = \pm A\,,~~(2y+1) b_1 A^2 = -6 \beta^2\,,
\ee
\be\label{3.45}
a_1 = [4(2-m)+6y]\beta^2 - b_1 A^2 y^2\,,~~
a_2 = (5-4m)\beta^2 - b_1 A^2 y^2\,,
\ee
while $y$ is as given by Eq. (\ref{2.66}).

There is a related PT-invariant solution with PT-eigenvalue $-1$
in one field and $+1$ in the other field. In particular, 
\bea\label{3.46}
&&\phi = A [\dn^2(\beta x, m)+y] +iD \sqrt{m} \dn(\beta x, m) \sn(\beta x, m)\,,
\nonumber \\
&&\psi = B \sqrt{m} \dn(\beta x, m) \sn(\beta x, m)
+iF [\dn^2(\beta x, m)+z]\,,
\eea
is an exact solution to Eqs. (\ref{3.1}), (\ref{3.2}) provided
\be\label{3.47}
b_1 = b_2 = \alpha\,,~~B= \pm A\,,~~F = \mp B\,,~~ y \ne z\,,~~
(y-z) b_1 A^2 y = -(3/2) \beta^2\,, 
\ee
\be\label{3.48}
a_1 = [5-4m+(3/2)(3y+z)]\beta^2\,,~~
a_2 = [5-4m+(3/2)(3z+y)]\beta^2\, , 
\ee
and both $y$ and $z$ are different; they are given by the two roots of Eq. (\ref{3.42b}).

There is another PT-invariant solution with PT-eigenvalue $-1$
in one field and $+1$ in the other field which is related to solution
(\ref{3.43}). In particular, 
\bea\label{3.46a}
&&\phi = A [\dn^2(\beta x, m)+y] +iD m \cn(\beta x, m) \sn(\beta x, m)\,,
\nonumber \\
&&\psi = B \sqrt{m} \dn(\beta x, m) \sn(\beta x, m)
+iF \sqrt{m} \cn(\beta x,m) \dn(\beta x, m)\,,
\eea
is an exact solution to Eqs. (\ref{3.1}), (\ref{3.2}) provided
\be\label{3.47a}
b_1 = b_2 = \alpha\,,~~B= \pm A\,,~~F = \mp B\,,~~ 
b_1 A^2 y = -(3/2) \beta^2\,,
\ee
\be\label{3.48a}
a_1 = [2-m+(9/2)y]\beta^2 -(1-m) b_1 A^2\,,~~
a_2 = [2-m+(3/2)y]\beta^2 -(1-m) b_1 A^2\, ,
\ee
and $y$ is given by Eq. (\ref{2.68}).

In the limit $m=1$, both the solutions (\ref{3.46}) and (\ref{3.46a}) go
over to the hyperbolic PT-invariant solution (\ref{3.39d}).

It is easy to check that
\be\label{3.50}
\phi = A \sqrt{m} \cn(\beta x, m) \dn(\beta x, m)\,,~~
\psi = B \sqrt{m} \sn(\beta x, m) \dn(\beta x, m)\,,
\ee
is an exact solution to the coupled Eqs. (\ref{3.1}), (\ref{3.2}) provided
\be\label{3.51}
b_1 = b_2 = \alpha\,,~~B = \pm A\,,~~m b_1 A^2 = -6 \beta^2\,,~~
a_1 = (5-m)\beta^2\,,~~a_2 = (5-4m)\beta^2\,.
\ee

There is a related PT-invariant solution with PT-eigenvalue $-1$
in one field and $+1$ in the other field. In particular, 
\bea\label{3.52}
&&\phi = A \sqrt{m} \cn(\beta x, m) \dn(\beta x, m) +iD m \cn(\beta x, m) 
\sn(\beta x, m)\,,
\nonumber \\
&&\psi = B \sqrt{m} \dn(\beta x, m) \sn(\beta x, m)
+iF [\dn^2(\beta x, m)+y]\,,
\eea
is an exact solution to Eqs. (\ref{3.1}), (\ref{3.2})  provided
\be\label{3.53}
b_1 = b_2 = \alpha\,,~~B= \pm A\,,~~F = \mp B\,,~~ y \ne z\,,~~
(y+1-m) b_1 A^2  = (3/2) \beta^2\,,
\ee
\be\label{3.54}
a_1 = (2-m)\beta^2 +[y^2-(1-m)]b_1 A^2\,,~~
a_2 = (5-4m+3y)\beta^2 +[y^2-(1-m)]b_1 A^2\,,
\ee
while $y$ is given by Eq. (\ref{3.42b}).

It is easy to check that
\be\label{3.55}
\phi = A \sqrt{m} \cn(\beta x, m) \dn(\beta x, m)\,,~~
\psi = B m \sn(\beta x, m) \cn(\beta x, m)\,,
\ee
is an exact solution to the coupled Eqs. (\ref{3.1}), (\ref{3.2}) provided
\be\label{3.56}
b_1 = b_2 = \alpha\,,~~B = \pm A\,,~~ b_1 A^2 = -6 \beta^2\,,~~
a_1 = (5m-1)\beta^2\,,~~a_2 = (5m-4)\beta^2\,.
\ee

There is a related PT-invariant solution with PT-eigenvalue $-1$
in one field and $+1$ in the other field. In particular, 
\bea\label{3.57}
&&\phi = A \sqrt{m} \cn(\beta x, m) \dn(\beta x, m) +iD \sqrt{m} 
\dn(\beta x, m) \sn(\beta x, m)\,,
\nonumber \\
&&\psi = B m \cn(\beta x, m) \sn(\beta x, m)
+iF [\dn^2(\beta x, m)+y]\,,
\eea
is an exact solution to Eqs. (\ref{3.1}), (\ref{3.2})  provided
\be\label{3.58}
b_1 = b_2 = \alpha\,,~~B= \pm A\,,~~F = \mp B\,,~~ y \ne z\,,~~
 y b_1 A^2  = (3/2) \beta^2\,,
\ee
\be\label{3.59}
a_1 = [(2-m)+(3/2)y]\beta^2 +(1-m)]b_1 A^2\,,~~
a_2 = [2-m+(9/2)]\beta^2 +(1-m)]b_1 A^2\,,
\ee
while $y$ is given by Eq. (\ref{2.68}).

In the limit $m=1$, both the solutions (\ref{3.52}) and (\ref{3.57}) go
over to the hyperbolic PT-invariant solution (\ref{3.39d}). 

It is easy to check that
\be\label{3.60}
\phi = A\left[\frac{(1-m)}{\dn^2(\beta x, m)}+y\right]\,,
~~\psi = B m \frac{\sn(\beta x, m)\cn(\beta x, m)}{\dn^2(\beta x, m)}\,,
\ee
is an exact solution to the coupled Eqs. (\ref{3.1}), (\ref{3.2}) provided
\bea\label{3.61}
&&b_1 = b_2 = \alpha\,,~~B = \pm A\,,~~(2y+2-m)b_1 A^2 = -6\beta^2\,,
\nonumber \\
&&a_1 = [4(2-m)+6y]\beta^2 -[y^2-(1-m)]b_1 A^2\,,~~
a_2 = (2-m)\beta^2 -[y^2-(1-m)]b_1 A^2\,,
\eea
while $y$ is given by Eq. (\ref{2.66}).

The PT-invariant combination 
\bea\label{3.62}
&&\phi = A  \left[\frac{(1-m)}{\dn^2(\beta x, m)}+y\right]
+iD m \frac{\cn(\beta x, m)\sn(\beta x, m)}{\dn^2(\beta x, m)} \nonumber \\
&&\psi = B m \frac{\cn(\beta x, m)\sn(\beta x, m)}{\dn^2(\beta x, m)}  
+iF \left[\frac{(1-m)}{\dn^2(\beta x, m)}+z\right]
\eea
is also an exact solution to Eqs. (\ref{3.1}), (\ref{3.2}) provided
\be\label{3.63}
b_1 = b_2 = \alpha\,,~~D = \pm A\,,~~F = \mp B\,,~~2 b_1 A^2 (y-z) 
= -3 \beta^2\,,
\ee
\be\label{3.65}
a_1 = [2-m+(3/2)(3y+z)]\beta^2\,,~~
a_2 = [2-m+(9/2)(y+3z)]\beta^2\,,
\ee
and $y$ and $z$ are unequal; they are given by the two roots of Eq. (\ref{2.68}).

\vskip 0.1truein 
\noindent{\bf Case III: Solutions With PT Eigenvalue $+1$ in Both The Fields}

Finally, let us discuss PT-invariant solutions in terms of Lam\'e polynomials
of order two with PT-eigenvalue $+1$ in both the fields. In this context we 
first note that 
\be\label{3.66}
\phi = A\left[\frac{(1-m)}{\dn^2(\beta x, m)}+y\right]\,,
~~\psi = B \sqrt{m} \frac{\cn(\beta x, m)}{\dn^2(\beta x, m)}\,,
\ee
is an exact solution to the coupled Eqs. (\ref{3.1}), (\ref{3.2}) provided
\be\label{3.67}
b_1 = b_2 = \alpha\,,~~B = \pm A\,,~~(2y+1)b_1 A^2 = -6\beta^2\,,~~
a_1 = [4(2-m)+6y]\beta^2 - b_1 A^2 y^2\,,~~
a_2 = (5-4m)\beta^2 - b_1 A^2 y^2\,,
\ee
while $y$ is given by Eq. (\ref{2.66}).

The PT-invariant combination 
\bea\label{3.68}
&&\phi = A  \left[\frac{(1-m)}{\dn^2(\beta x, m)}+y\right]
+iD m \frac{\cn(\beta x, m)\sn(\beta x, m)}{\dn^2(\beta x, m)} \nonumber \\
&&\psi = B \sqrt{m} \frac{\cn(\beta x, m)}{\dn^2(\beta x, m)}  
+iF \sqrt{m(1-m)}\frac{\sn(\beta x, m)}{\dn^2(\beta x, m)}
\eea
is also an exact solution to Eqs. (\ref{3.1}), (\ref{3.2}) provided
\be\label{3.69}
b_1 = b_2 = \alpha\,,~~D = \pm A\,,~~F = \mp B\,,~~2 b_1 A^2  
= -3 \beta^2\,,
\ee
\be\label{3.71}
a_1 = [2-m+(9/2)y]\beta^2 -(1-m) b_1 A^2\,,~~
a_2 = [2-m+(3/2)y]\beta^2 -(1-m) b_1 A^2\,,
\ee
while $y$ is given by Eq. (\ref{2.68}).

\subsection{Coupled KdV Equations}

We now consider a coupled KdV model 
\cite{zhou} which has received some attention in the literature. In our
recent paper \cite{ks}, we have already obtained two PT-invariant solutions with
PT-eigenvalue $+1$ in both the fields. We now show that the same model has
another novel PT-invariant periodic solution with PT-eigenvalue $+1$ in 
both the fields.   

The coupled KdV equations are
\bea\label{4.1}
&&u_t + \alpha u u_x + \eta v v_{x} + u_{xxx} = 0\,, \nonumber \\
&&v_t + \delta u v_x + v_{xxx} = 0\,.
\eea
It is easy to check that the coupled Eqs. (\ref{4.1}) have the periodic
solution
\be\label{4.2}
u = \frac{A(1-m)}{\dn^2[\beta(x-ct), m]}\,,~~
v = \frac{B(1-m)}{\dn^2[\beta(x-ct), m]}\,,
\ee
provided
\be\label{4.3}
\delta A = 12 \beta^2\,,~~\eta B^2 = (\delta - \alpha)A^2\,,~~
c = 4(2-m) \beta^2\,.
\ee
Remarkably, the same model also admits the PT-invariant periodic solution
\bea\label{4.4}
&&u = \frac{A(1-m)}{\dn^2[\beta(x-ct),m]}
+iD m\sqrt{1-m} \frac{\sn[\beta(x-ct),m]\cn[\beta(x-ct),m]}
{\dn^2[\beta(x-ct), m]}\,, \nonumber \\
&&v = \frac{B(1-m)}{\dn^2[\beta(x-ct), m]}
+iF m\sqrt{1-m} \frac{\sn[\beta(x-ct),m]\cn[\beta(x-ct),m]}
{\dn^2[\beta(x-ct), m]}\,, 
\eea
provided
\be\label{4.5}
D = \pm A\,,~~F = \pm B\,,~~\delta A = 6 \beta^2\,,~~\eta B^2 
= (\delta - \alpha)A^2\,,~~c = (2-m) \beta^2\,.
\ee
Note that the signs of $D = \pm A$ and $F = \pm B$ are correlated.
It is worth pointing out that this nonsingular solution vanishes
in the hyperbolic limit of $m=1$.

\subsection{Coupled KdV-mKdV Model}

Recently we had considered \cite{ks} a coupled KdV-mKdV model \cite{coupled} 
\bea\label{5.1}
&&u_t + u_{xxx} + 6u u_x+2\alpha u v v_x = 0\,, \nonumber \\
&&v_t + v_{xxx} + 6v^2 v_x+\gamma  v u_x = 0\,,
\eea
and obtained PT-invariant solutions with PT-eigenvalue $+1$ in both the fields.
We now show that the same model also admits PT-invariant solutions with
PT-eigenvalue $+1$ in one field and $-1$ in the other field.

Let us first note that
\be\label{5.2}
u = A\dn^2[\beta(x-ct),m]+A y\,,~~v = B\sqrt{m} \sn[\beta(x-ct),m]\,,
\ee
is an exact solution of the coupled Eqs. (\ref{5.1}) provided
\be\label{5.3}
4\gamma A -12 B^2 = 12 \beta^2 = 6A -\alpha B^2\,,~~
c = -(1+m) \beta^2\,,~~ y = -\frac{(3-m)}{4}\,.
\ee
It is easy to check that the same model also admits the PT-invariant solution
\bea\label{5.4}
&&u = A[\dn^2[\beta(x-ct),m]+y]+iD \sqrt{m} \sn[\beta(x-ct),m]\dn[\beta(x-ct),m]
\,, \nonumber \\
&&v = B\sqrt{m} \sn[\beta(x-ct),m]+iF \dn[\beta(x-ct),m]\,,
\eea
provided
\be\label{5.5}
D = \pm A\,,~~F = \mp B\,,~~2\gamma A - 12 B^2 = 3 \beta^2 = 3A -\alpha B^2\,,~~
c = -\frac{(2m-1)}{2} \beta^2\,,~~y = -\frac{(3-2m)}{4}\,.
\ee
Note that the signs of $D = \pm A$ and $F = \mp B$ are correlated.
Further, the same model also admits another PT-invariant solution
\bea\label{5.6}
&&u = A[\dn^2[\beta(x-ct),m]+ y] +iD m \sn[\beta(x-ct),m]\cn[\beta(x-ct),m]\,,
\nonumber \\
&&v = B \sqrt{m} \sn[\beta(x-ct),m] +iF \sqrt{m} \sn[\beta(x-ct),m]\,,
\eea
provided
\be\label{5.7}
D = \pm A\,,~~F = \mp B\,,~~2\gamma A -12B^2 = 3 \beta^2 = 3A -\alpha B^2\,,~~
c = -\frac{(2-m)}{2} \beta^2\,,~~G = -\frac{(2-m)}{4}\,.
\ee
Note that the signs of $D = \pm A$ and $F = \mp B$ are correlated.

In the limit $m=1$, both the solutions (\ref{5.4}) and (\ref{5.6}) go over
to the hyperbolic mixed PT-invariant solution
\bea\label{5.4a}
&&u = A[\sech^2 \beta(x-ct)+y]+iD \sech \beta(x-ct)\tanh \beta(x-ct)
\,, \nonumber \\
&&v = B \tanh \beta(x-ct)+iF \sech \beta(x-ct)\,,
\eea
provided 
\be\label{5.5a}
D = \pm A\,,~~F = \mp B\,,~~2\gamma A - 12 B^2 = 3 \beta^2 
= 3A -\alpha B^2\,,~~c = -1/2 \beta^2\,,~~y = -1/4\,.
\ee
Note that the signs of $D = \pm A$ and $F = \mp B$ are correlated.

Yet another solution to the coupled Eqs. (\ref{5.1}) is
\be\label{5.8}
u = \frac{A(1-m)}{\dn^2[\beta(x-ct),m]}+A y\,,~~v = B\sqrt{m(1-m)} 
\frac{\sn[\beta(x-ct),m]}{\dn[\beta(x-ct), m}\,,
\ee
provided
\be\label{5.9}
4\gamma A +12 B^2 = 12 \beta^2 = 6A +\alpha B^2\,,~~
c = (2m-1) \beta^2\,,~~ y = -\frac{(3-2m)}{4}\,.
\ee
It is easy to check that the same model also admits the PT-invariant solution
\bea\label{5.10}
&&u = \frac{A(1-m)}{[\dn^2[\beta(x-ct),m]}+y+iD \sqrt{m(1-m)} 
\frac{\cn[\beta(x-ct), m]\sn[\beta(x-ct),m]}{\dn^2[\beta(x-ct),m]}
\,, \nonumber \\
&&v = B\sqrt{m(1-m)} \frac{\sn[\beta(x-ct),m]}{\dn[\beta(x-ct), m]}
+iF\sqrt{m}\frac{\cn[\beta(x-ct), m]}{\dn[\beta(x-ct),m]}\,,
\eea
with mixed PT-eigenvalues provided
\be\label{5.11}
D = \pm A\,,~~F = \pm B\,,~~2\gamma A - 12 B^2 = 3 \beta^2 = 3A -\alpha B^2\,,~~
c = -\frac{(2-m)}{2} \beta^2\,,~~y = -\frac{(2-m)}{4}\,.
\ee
Note that the signs of $D = \pm A$ and $F = \pm B$ are correlated.

\section{Summary and Conclusions}

In this paper we have in a sense completed the program which we had 
initiated recently. In particular, in a recent publication \cite{ks} we have 
shown through several examples that whenever a
real nonlinear equation admits solutions in terms of $\sech x$ $(\sech^2 x$), 
then the same equation also admits complex PT-invariant solutions with 
PT-eigenvalue $+1$ in terms of $\sech x \pm i \tanh x$ 
$(\sech^2 x \pm i \sech x \tanh x$). Further, we have also shown that such
PT-invariant solutions also exist in the corresponding periodic case. 
On the other hand, in this paper we have shown through several examples
 that whenever a real nonlinear equation admits kink solutions in terms of 
$\tanh x$, then the same equation also admits complex PT-invariant kink 
solutions with PT-eigenvalue $-1$ in terms of $\tanh x \pm i \sech x$.
We have also shown that both the kink and the PT-invariant kink solutions
have the same topological charge as well as the same kink energy. 
In addition, for several kink bearing equations we have explicitly 
shown that even the PT-invariant kink solution is linearly stable. In
the case of $\phi^4$ kink we have shown that
like the usual $\phi^4$ kink, the PT-invariant $\phi^4$ kink
too has two modes and the corresponding eigenenergies are in fact
identical for the usual and the PT-invariant kink. We believe this is quite 
significant and the PT-invariant kink can have some physical relevance. It 
would be worthwhile to examine this issue in detail.

Further, we have shown that such PT-invariant solutions also exist in the 
corresponding periodic case.  In particular, we have shown through several 
examples that whenever a nonlinear equation admits
periodic solutions in terms of Jacobi elliptic function $\sn(x,m)$, then
the same equation will also admit complex PT-invariant periodic solutions
with PT-eigenvalue $-1$ in terms of $\sn(x,m) \pm i \cn(x,m)$ as well as
$\sn(x,m) \pm i \dn(x,m)$. Moreover, in a few coupled models we have also 
shown the existence of PT-invariant periodic solutions in terms of Lam\'e
polynomials of order one and two and with PT-eigenvalue being either
$+1$ or $-1$ in both the fields or $+1$ in one field and $-1$ in the other field. 

These results raise several important questions. The obvious open question 
is whether these result are true in general. It
would be nice if one can prove this in general, both in the hyperbolic 
as well as in the periodic case. In the absence of a general proof, 
it is worthwhile to look at more examples and see if this 
observation is true in the additional cases or if there are some 
exceptions. The other obvious question is:  
what could be the deeper reason because of which such solutions
exist? Another question is about the significance of
such solutions for real nonlinear equations. In this context we would like
to remark that the symmetry of solutions of a nonlinear equation need not
be the same as that of the nonlinear equation but it could be less. 
We believe that auto-B\"acklund transformations may also provide the 
solutions considered here in the case of both integrable \cite{backlund} 
and non-integrable models \cite{back2}. 

Normally, the complex solutions of a real nonlinear equation are not of 
relevance. However, being PT-invariant complex solutions, we believe 
they could have some physical significance including in coupled optical 
waveguides \cite{book,peng, ruter}. 
One pointer in this direction is the fact that 
for both the KdV and the mKdV equations, which are integrable equations, we 
have checked that the first three constants of motion for the PT-invariant complex 
solutions of both the KdV and the mKdV equations are in fact real but have 
different values than those for the usual hyperbolic solution (and we suspect 
that in fact all the constants of motion would be real and would be different 
than those for the real hyperbolic solution) thereby suggesting that such 
solutions could be physically interesting. Thus, it would be worthwhile to 
study the stability of such PT-invariant solutions, which may shed some light 
on the possible significance of these solutions.  We hope to address some of 
these issues in the near future.

\section{Acknowledgments} 

One of us (AK) is grateful to Indian National Science Academy (INSA) 
for the award of INSA Senior Scientist position at 
Savitribai Phule Pune University. This work was supported in part by the 
U.S. Department of Energy.

\end{document}